\documentclass[12pt,thmsa]{article}

\begin{document}

\begin{center}
{\LARGE \textbf{$n=3$ Differential calculus and gauge theory on a reduced
quantum plane.}}\\[0pt]\vspace{1cm}

\vspace{0.5cm}

\vspace{0.5cm}

{\large M. EL BAZ\footnote{{\large moreagl@yahoo.co.uk}}, A. EL HASSOUNI
\footnote{{\large lhassoun@fsr.ac.ma}}, Y. HASSOUNI\footnote{{\large %
y-hassou@fsr.ac.ma}}\\ \ and E.H. ZAKKARI\footnote{{\large %
hzakkari@hotmail.com}}.}\\[0pt]\vspace{0.5cm}\vspace{0.5cm}

Laboratory of Theoretical Physics\\[0pt]PO BOX 1014, University Mohammed V %
\\[0pt]Rabat, Morocco.
\end{center}

\bigskip

\vspace {2cm}

\textbf{Abstract:} We discuss the algebra of $N\times N$ matrices as a
reduced quantum plane. \ A $3-$nilpotent deformed differential calculus
involving a complex parameter $q$ is constructed. The two cases, $q$ $3^{rd}$
and $N^{th}$ root of unity are completely treated. As application, a gauge \
field theory for the particular cases $n=2$ and $n=3$ is established.\

\vspace {2cm}


\newpage\

\section{Introduction:}

An adequate way leading to generalizations of the ordinary exterior
differential calculus arises from the graded differential algebra $\left[
1-3\right] $. These generalizations are not universal as far as we know, and
many technics have been used to introduce differential calculus that
corresponds to non commutative calculi. The latters involve a complex
parameter that satisfies some conditions allowing the obtention of a
consistent generalized differential calculus. It is usually called $q$%
-differential calculus. In ref $\left[ 1-3\right] $ , it is seen
as a graded \ $q$-differential algebra which is the sum of $\
k-$graded subspaces, where $k\in \left\{ 0,1,\, 2...m-1\right\} .$
The relevent differential operator is an endomorphism $d$ \
satisfying $d^m=0$ and the $q$-Leibniz rule:

$\ \ \ \ \ \ \ \ \ \ \ \ \ \ \ \ \ \ \ \ \ \ \ \ \ \ \ \ \ \ \ \ \ \
d(AB)=(dA)B+qAd(B).$

The most important property of this calculus is that it contains not only
first differentials $dx^i,$ $i=1...n,$ but also it involves the higher-order
differentials $d^jx^i,j=1....m-1$.

On the other hand, the differential calculi $(d^2=0)$ on noncommutative
spaces was also studied by different authors, see for exemple $[4-9].$ The
common property of these calculi is the covariance of these latters under
some symmetry quantum group.

In this paper, we construct a \ covariant differential calculus \ $d^3=0$ on
the algebra $M$ of $3\times 3$ matrices considered as a quantum plane. We
will show that our differential calculus is covariant under the algebra of
transformations with a quantum group structure. The complex deformation
parameter $q$ ($3^{rd}-$root of unity) will play an important role in
constructing the differential calculus that we introduce. As it is done in
the litterature of deformed differential calculus $\left[ 4,5\right] $, this
case implies a non trivial study. As application, we treat the gauge field
theory.

The paper is organized as follows:

We start in section $2$ by defining the algebra of $N\times N$ matrix \ as a
reduced quantum plane, where the deformation parameter $q$ is $N$-th root of
unity. We also give a matrix realization in the case $N=3$. In section $3$
we construct the covariant differential calculus $d^3=0,$ on two dimensional
reduced quantum plane as in ref $\left[ 1-3\right] $. The new objects, $d^2x$
and $d^2y,$ appearing in this construction are seen as the analogous of the
differential elements $dx$ and $dy$ in the ordinary differential calculus.
In section $4,$ we generalize this result by considering a complex
deformation parameter $q$ $N^{th}$ root of unity.

In section $5$, we study the application of this new differential calculus $%
(N=3)$ to the gauge field theory on $M_3(C).$ We recall in section $6$ the
differential calculus $d^2=0$ $\left[ 6-9\right] $, and we apply it to the
gauge theory on $M_3(C)$.

\section{Preliminaries about the algebra \ M$_3$(C) of N$\times $N matrices
as a reduced quantum plane.}

The associative algebra of $N\times N$ \ matrices is generated by two
elements $x$ and \ $y$ $[10]$ satisfying the relations:

\begin{equation}
xy=qyx
\end{equation}

and

\begin{equation}
x^N=y^N=1,
\end{equation}

where $1$ is the unit matrix and $q$ $(q\neq 1)$ is a complex parameter $%
N^{th}$ root of unity.

In the case $N=3$, an explicit matrix realization of generators $x$ and $y$ $%
\left[ 6,11\right] $ \ is given by:

\begin{equation}
x=\left(
\begin{array}{ccc}
1 & 0 & 0 \\
0 & q^{-1} & 0 \\
0 & 0 & q^{-2}
\end{array}
\right)
\end{equation}

\begin{equation}
y=\left(
\begin{array}{ccc}
0 & 1 & 0 \\
0 & 0 & 1 \\
1 & 0 & 0
\end{array}
\right) ,
\end{equation}

and $q$ satisfies the relation:

\begin{equation}
1+q+q^{2}=0.
\end{equation}

The associative algebra, noted by $C_q$ $\left[ x,y\right] :=C_q,$ of formal
power series defined over the two dimentional quantum plane is generated by $%
x$ and $y$ with a single quadratic relation $xy=$ $qyx$. it is clear that $%
C_1$ $\left[ x,y\right] $ coincides with the algebra of polynomials over
commuting variables $x,$ $y.$

Note that if the generators $x,$ $y$ do not satisfy any additional
relations, then $C_q$ is infinite dimensional. In the case of the algebra $%
M_3(C)$ of $3\times 3$ matrices over complex numbers, the generators $x,$ $y$
satisfy the above quadratic relation and the cubic ones $x^3=y^3=1$. Thus it
is generated by the following set: \{$1,$ $x,$ $y,$ $x^2,$ $y^2,$ $xy,$ $%
x^2y,$ $xy^2,$ $x^2y^2$\}. In this case, the algebra $M_3(C)$ appears as the
associative quotient algebra $\ C_q^0$ by the bilateral ideal generated by $%
x^3$-$1=0$ and $y^3-1=0$. Here $C_q^0$ is the unital extension of $C_q.$
That is, in the sense of $\left[ 6,11\right] $, the $3\times 3$ matrices
over $C$ are seen as a reduced quantum plane. \

We note that the functions of $x$ and $y$ are seen as formal power series
with a maximum degree $3;$ this property will be extremely useful in what
follows. In fact, the set of those functions is an associative algebra that
is used to introduce a gauge field theory on the reduced quantum plane. This
idea will be developed in sections $5$ and $7$.

\section{Differential calculus with nilpotency $n=3$ on reduced quantum
plane, case $q^3=1$}

The aim of this section is to construct a covariant $n=3$ nilpotent
differential calculus by mixing two approaches; namely we adapt to the
reduced quantum plane an idea originally proposed by Kerner $[1-3]$, and we
use Couquereaux's technics $[6]$ to ensure covariance. We denote by $\Omega $
the differential algebra generated by $\ x$, $y$, $dx$, $dy$, $d^2x$ and $%
d^2y$, where the ''2- forms'' $d^2x$ and $d^2y$ are the second differentials
of the basic variables $x$ and $y$.

Let us introduce the differential operator $d$ that satisfies the following
conditions:

Nilpotency,

\begin{equation}
d^3=0.
\end{equation}

Leibniz rule,

\begin{equation}
d(uv)=d(u)v+q^nud(v),
\end{equation}

where $u$ is a form of degree $n$ and $q$ is $3^{rd}$ root of unity.

By applying the Leibniz rule on the $1$--form we obtain:

\begin{equation}
d(f(x)\, dx)=(df(x))\, dx+f(x)\, d^2x,
\end{equation}
$f(x)$ are the $0-$form in the algebra $\ \Omega $. The notion of covariance
is necessary for the consistency of every differential calculus. The set of
transformations leaving covariant our differential calculus is $F\subset
Fun(SL_q(2,C))$ and the covariance is described by the left coaction. We
start by explaining this coaction $\left[ 12\right] .$

The left coaction of the group $F$ on the reduced quantum plane is the
linear transformation of coordinates given by:

\[
\left(
\begin{array}{c}
x_{1} \\
y_{1}
\end{array}
\right) =\left(
\begin{array}{c}
a\, b \\ c\, d
\end{array}
\right) \otimes \left(
\begin{array}{c}
x \\
y
\end{array}
\right) .
\]

We introduce also the line vectors with coordinate functions:

\[
\left( x^1,\, y^1\right) =\left( x,\, y\right) \otimes \left(
\begin{array}{c}
a\, b \\ c\, d
\end{array}
\right) ,\ \ \
\]

where the matrix elements $a,$ $b,$ $c$ and $d$ do not commute with each
others.We require that the quantities $x_1,$ $y_1,$ $x^1,$ $y$\ $^1$
obtained in \ the above relations satisfy the same relations as $x$ and $y.$
The two constraints $x_1y_1=qy_1x_1$ and $x^1y^1=qy^1x^1$ lead to the
relations:

$\ \ \ \ \ \ \ \ \ \ \ \ \ \ \ \ \ \ \ \ \ \ \ \ \ \ \ \ \ \ \ \ \ \ \ \ \ \
\ \ \ \ \ \ \ \ \ \ \ \ \ \ \ \ \ \ \ \ \ \ \ \ \ \ \ \ \ \ \ \ \ \ \ \ $%
\[
\ \ \ \ \ \ \ ac=qca\ \ \ \ \ \ \ \ \ \ \ \ \ \ \ bd=qdb
\]

$\ \ \ \ \ \ \ \ \ \ \ \ \ \ \ \ \ \ \ \ \ \ \ \ \ \ \ \ \ \ \ \ \ \ \ \ \ \
\ \ \ \ \ \ \ \ \ \ \ \ \ \ \ \ \ \ \ \ \ \ \ \ \ \ \ \ \ \ \ \ \ \ \ $%
\[
\ \ \ \ \ \ \ \ ab=qba\ \ \ \ \ \ \ \ \ \ \ \ \ \ \ \ cd=qdc
\]

$\ \ \ \ \ \ \ \ \ \ \ \ \ \ \ \ \ \ \ \ \ \ \ \ \ \ \ \ \ \ \ \ \ \ \ \ \ \
\ \ \ \ \ \ \ \ \ \ \ \ \ \ \ \ \ \ \ \ \ \ \ \ \ \ \ \ \ \ \ \ \ $%
\[
\ \ \ \ \ \ \ \ \ \ \ \ bc=cb\ \ \ \ \ \ \ \ \ \ \ \ \ \ ad-da=(q-q^{-1})bc.
\]

The algebra generated by $a,\ b,\ c,$ and $d$ is usually denoted $%
Fun(GL_q(2,C)).$ The $q$-determinant $D=da-q^{-1}bc$ is in the center of $%
Fun(GL_q(2,C)).$ If we set it to be equal to $1,$ we define the algebra $%
Fun(SL_q(2,C)).$ Assuming that the supplementary conditions $(x)^3=1$ and $%
(y)^3=1$ are also verified by the coordinates $x_1$, $\ y_1$ $($and $\ x^1$,
$y^1)$, $($ $(x_1)^3=(y_1)^3=1$, $((x^1)^3=(y^1)^3=1)),$ implies $\ a^3=1,$ $%
b^3=0,$ $c^3=0,$ $d^3=1$. These new cubic relations on $Fun(SL_q(2,C)),$
yields a new algebra that we denote $F$. It is also a Hopf algebra. Indeed,
it has a coalgebra structure (coproduct) which is compatible with the
algebra one (product), this defines a bialgebra structure. An antipode and a
co-unit are ~also defined . For further details on such structures on $F$,
see, for example $[6]$.

The mixture of Kerner's idea and Coquereaux's technics allows us to
construct the left covariant differential algebra $\Omega =\{x,$ $y,$ $dx,$ $%
dy,$ $d^2x,$ $d^2y\},$ see appendix$.$ The commutation relations between the
generators of $\Omega $ are as follows:

\begin{equation}
x\, dx=q^{2}dx\, x
\end{equation}

\begin{equation}
x\, dy=qdy\, x+(q^2-1)dx\, y
\end{equation}

\begin{equation}
y\, dx=qdx\, y
\end{equation}

\begin{equation}
y\, dy=q^2dy\, y
\end{equation}

\begin{equation}
dy\, dx=q^2dx\, dy
\end{equation}

\begin{equation}
x\, d^2x=q^2d^2x\, x
\end{equation}

\begin{equation}
y\, d^2x=qd^2x\, y
\end{equation}

\begin{equation}
y\, d^2y=q^2d^2y\, y
\end{equation}

\begin{equation}
x\, d^2y=qd^2y\, x+(q^2-1)d^2x\, y
\end{equation}

\begin{equation}
dx\, d^2y=d^2y\, dx+q(1-q)d^2x\, dy
\end{equation}

\begin{equation}
dy\, d^2x=d^2x\, dy
\end{equation}

\begin{equation}
dx\, d^2x=qd^2x\, dx
\end{equation}

\begin{equation}
dy\, d^2y=qd^2y\, dy
\end{equation}

\begin{equation}
d^{2}y\, d^{2}x=q^{2}d^{2}x\, d^{2}y.
\end{equation}

As in the standard way, we define the partial derivatives in directions $x$
and $y$ through:

\begin{equation}
d=\frac \partial {\partial x}dx+\frac \partial {\partial y}dy=\partial
_xdx+\partial _ydy.
\end{equation}

Consistency conditions as in $[9]$ yield:

\begin{equation}
\partial _{x}\, \partial _{y}=q\partial _{y}\, \partial _{x}
\end{equation}

\begin{equation}
\partial _{x}\, x=1+q^{2}x\, \partial _{x}+(q^{2}-1)y\, %
\partial _{y}
\end{equation}

\begin{equation}
\partial _{x}\, y=qy\, \partial _{x}
\end{equation}

\begin{equation}
\partial _{y}\, y=1+q^{2}y\, \partial _{y}
\end{equation}

\begin{equation}
\partial _{y}\, y=1+q^{2}y\, \partial _{y}
\end{equation}

\begin{equation}
(dx)^{3}=(dy)^{3}=0.
\end{equation}

The last equality $eq(29)$ can be related to the nilpotency relation
encountered \ in the description of the fractional statistics. More
precisely, we recover the description of physical systems that generalize
fermions. In a forthcoming paper $\left[ 13\right] $, we reintroduce these
systems using this new differential calculus by establishing an adequate
correspondence between our differential calculus and some deformed
Heinsenberg algebras, as it is done in $\left[ 14\right] $ for the
particular case $(d^2=0)$.

Now, we generalize our differential calculus by considering the case $q^N=1$.

\section{Differential calculus on a reduced quantum\protect\\plane, case $%
q^N=1$.}

A two dimensional reduced quantum plane is an associative algebra generated
by $x$ and $y$ with relations $(1)$ and $(2).$ One can always define the
differential operator ''$d$'' satisfying $d^3=0,$ $(d^2\neq 0),$ and the
Leibniz rule:

\begin{equation}
d(uv)=d(u)v+(j)^nud(v),
\end{equation}

$u\in \Omega ^n$ and $v\in \Omega ^m$, where $\Omega ^n$ and $\Omega ^m$ are
the spaces of $n$ and $m$ forms on reduced quantum plane respectively.

Note that in contrast to $eq(7),$ one have to distinguish between the
deformation parameter $q$ and the $j$ parameter, $j^3=1$, $(j\neq 1)$, in ~$%
eq(30)$.

Following the same method of section $3,$ we get the covariant differential
calculus:

\begin{equation}
x\, dx=j^{2}dx\, x
\end{equation}

\begin{equation}
x\, dy=-\frac{jq}{1+q^2}dy\, x+\frac{j^2q^2-1}{1+q^2}dx\, y
\end{equation}

\begin{equation}
y\, dx=\frac{j^2-q^2}{1+q^2}dy\, x-\frac{jq}{1+q^2}dx\, y
\end{equation}

\begin{equation}
y\, dy=j^2dy\, y
\end{equation}

\begin{equation}
dx\, dy=qdy\, dx
\end{equation}

\begin{equation}
x\, d^{2}x=j^{2}d^{2}x\, x
\end{equation}

\begin{equation}
x\, d^{2}y=-\frac{jq}{1+q^{2}}d^{2}y\, x+\frac{j^{2}q^{2}-1}{%
1+q^{2}}d^{2}x\, y
\end{equation}

\begin{equation}
yd^2x=\frac{j^2-q}{1+q^2}^2d^2y\, x-\frac{jq}{1+q^2}d^2x\, y
\end{equation}

\begin{equation}
y\, d^2y=j^2d^2y\, y
\end{equation}

\begin{equation}
dx\, d^{2}x=jd^{2}x\, dx
\end{equation}

\begin{equation}
dx\, d^{2}y=-\frac{q}{1+q^{2}}d^{2}y\, dx+\frac{jq^{2}-j^{2}}{%
1+q^{2}}d^{2}x\, dy
\end{equation}

\begin{equation}
dyd^{2}x=\frac{j-j^{2}q}{1+q^{2}}^{2}d^{2}y\, dx-\frac{q}{1+q^{2}}d^{2}x%
\, dy
\end{equation}

\begin{equation}
dy\, d^{2}y=jd^{2}y\, dy
\end{equation}

\begin{equation}
d^2x\, d^2y=qd^2y\, d^2x.
\end{equation}

We recover the differential calculus obtained in section $3,$ if $q=j$. As
an application of this new differential calculus $d^3$ $=0$ on the reduced
quantum plane, we construct in the section below a gauge field theory on $%
M_3(C).$

\section{Gauge theory on $M_3(C)$ as a reduced quantum plane with $d^3=0$}

In this section, we use the $n=3$ differential calculus constructed in
section $3$ to establish a gauge theory on the reduced quantum plane.

As in the ordinary case, the covariant differential is defined by:

\begin{equation}
D\Phi (x,y)=d\Phi (x,y)+A(x,y)\Phi (x,y),
\end{equation}
where the field $\Phi (x,y)$ is a function on $M_3(C)$ and the gauge field $%
A(x,y)$ is a $1$-form valued in the associative algebra of functions on the
reduced quantum plane $M_3(C)$.

We have assumed that the algebra of functions on $M_3(C)$ is a bimodule over
the differential algebra $\Omega .$

As usual, the covariant differential $D$ must satisfy:

\begin{equation}
DU^{-1}\Phi (x,y)=U^{-1}D\Phi (x,y),
\end{equation}

where $U$ is an endomorphism defined on $Fun[M_3(C)]$ .

This leads to the following gauge field transformation:

\begin{equation}
A(x,y)\rightarrow U^{-1}A(x,y)U+U^{-1}dU.
\end{equation}

In general, the 1-form gauge field $A(x,y)$ can be written as:

\begin{equation}
A(x,y)=A_{x}(x,y)dx+A_{y}(x,y)dy.
\end{equation}

The differential calculus $n=3$ allows to define the curvature $R$ as
follows $\left[ 2,15\right] $:

\begin{equation}
D^{3}\Phi (x,y)=R\Phi (x,y).
\end{equation}

Direct computations show that $R$ is a ''three-form'' given by:

\begin{eqnarray}
R&=& d^2A(x,y)+dA^2(x,y)+A(x,y)dA(x,y)+A^3(x,y) \\
&=& d^2A(x,y)+(dA(x,y))A(x,y)+(1+q)A(x,y)dA(x,y)+A^3(x,y) \\
&=& d^{2}A(x,y)+(dA(x,y))A(x,y)-q^{2}A(x,y)dA(x,y)+A^{3}(x,y).
\end{eqnarray}

One has to express the curvature written above in terms of $3$-forms
constructed from basic generators $dx,$ $dy,$ $d^2x$ and $d^2y$ \ of the
differential algebra $\Omega .$ Since we are dealing with a non-commutative
space (reduced quantum plane), this task is not straightforward. In fact,
the non-commutativity prevents us from rearanging the different terms in $%
eq(52)$ adequately. To overcome this technical difficulty we require that
the components of the gauge field $A_x(x,y)$ and $A_y(x,y)$ are expressed as
formal power series of the space coordinates $x$ and $y$ $\left[
16-19\right] $. The condition $eq(2)$ in section$2$ $(N=3)$ is extremely
useful, in the sense that it limits the power series to finite ones rather
than infinite:

\begin{equation}
A_{x}(x,y)=a_{mn}x^{m}y^{n};\, m,n=0,1,2
\end{equation}

\begin{equation}
A_y(x,y)=b_{kl}x^ky^l;\, k,l=0,1,2.
\end{equation}

Using the formulae $(1,31-44,52-54),$ and after technical computations, the
desired expression of the curvature arises as:

$R=\Big[ R_{xxy}+qR_{yxx}+q^2R_{xyx}+$

$(1-q)\Big\{ \partial _yA_x(x,y)+q\partial _xA_y(x,y)+\partial
_yA_Y(x,y)((1-q)f_2(y)-f_1(x,y))$

$+qf_4(x,y)f_0(x,y)-q^2f_6(x,y))+A_y(x,y)(f_5(x,y)+$

$A_y(x,y)A_y(q^2x,y)((1-q)f_2(y)-f_1(x,y))A_y(x,y)A_x(q^2x,y)f_4(x,y)+$

$qA_x(x,y)A_y(qx,q^2y)f_0(x,y)+q^2A_y(x,y)f_4(x,y)A_y(x,y)+$

$A_y(x,y)f_3(x,y)A_y(q^2x,qy) \Big\} \Big] dxdxdy$

$+\Big[ R_{yyx}+qR_{yxy}+q^2R_{xyy}+(1-q) \Big\{ -q^2\partial
_yA_y(x,y)f_0(x,y)-q^2A_y(x,y)f_7(x,y)-$

$A_y(x,y)A_y(q^2x,qy)f_8(x,y)\Big\} \Big] %
dydydx+qF_{xy}^qd^2xdy+F_{yx}^{q^2}d^2ydx,$

\vspace{0.5cm}

where:
\vspace{0.5cm}

$R_{xxy}=\partial _x\partial _xA_y(x,y)+\partial
_xA_x(x,y)A_y(q^2x,qy)-q^2A_x(x,y)\partial _xA_y(qx,q^2y)+$

\[
A_x(x,y)A_x(qx,q^2y)A_y(q^2x,qy)
\]

$R_{yxx}=\partial _y\partial _xA_x(x,y)+\partial
_yA_x(x,y)A_x(x,y)-q^2A_y(x,y)\partial _xA_x(qx,q^2y)+$

\[
A_y(x,y)A_x(q^2x,qy)A_x(x,y)
\]

$R_{xyx}=\partial _x\partial _yA_x(x,y)+\partial
_xA_y(x,y)A_x(x,y)-q^2A_x(x,y)\partial _yA(qx,q^2y)+$

\begin{equation}
A_x(x,y)A_y(qx,q^2y)A_x(x,y)
\end{equation}

$R_{yyx}=\partial _y\partial _yA_x(x,y)+\partial
_xA_y(x,y)A(x,y)-q^2A_x(x,y)\partial _yA_x(qx,q^2y)+$

\[
A_y(x,y)A_y(q^2x,qy)A_x(qx,q^2y)
\]

$R_{yxy}=\partial _y\partial _xA_y(x,y)+\partial
_yA_x(x,y)A_x(x,y)-q^2A_y(x,y)\partial _xA_y(qx,q^2y)+$

\[
A_y(x,y)A_x(q^2x,qy)A_y(x,y)
\]

$R_{xyy}=\partial _x\partial _yA_y(x,y)+\partial
_xA_y(x,y)A_y(x,y)-q^2A_x(x,y)\partial _yA_y(qx,q^2y)+$

\[
A_x(x,y)A_y(qx,q^2y)A_y(x,y)
\]

\begin{equation}
\begin{array}{ccl}
F_{xy}^q &=&\partial _xA_y(x,y)-q\partial _yA_x(x,y)+
A_x(x,y)A_y(qx,q^2y)-qA_y(x,y)A_x(q^2x,qy) \cr
F_{yx}^{q^2}&=&\partial _yA_x(x,y)-q^2\partial _xA_y(x,y)+
A_y(x,y)A_x(q^2x,qy)-q^2A(x,y)A(qx,q^2y)  \cr
\end{array}
\end{equation}

\begin{eqnarray}
f_0(x,y) & =& -b_{11}y^2-qb_{10}y+q^2b_{22}x+b_{20}xy+qb_{21}xy^2-b_{21}
\nonumber \\
f_1(x,y)&=& -a_{11}y^2-a_{10}y+a_{22}x+a_{20}xy+a_{21}xy^2-a_{12}  \nonumber
\\
f_2(x,y)&=& -b_{20}y^2-q^2b_{21}y-q^2b_{22}y  \nonumber \\
f_3(x,y) & =& -q^2a_{11}y^2-a_{10}y+q^2a_{22}x+qa_{20}xy+a_{21}xy^2-qa_{12}
\nonumber \\
f_4(x,y) & =& -q^2b_{11}y^2-b_{10}y+q^2b_{22}x+qb_{20}xy+b_{21}xy^2-qb_{12}
\\
f_5(x,y) &=& -qb_{21}y^2-b_{20}y-qb_{22}  \nonumber \\
f_6(x,y) &=& +qa_{12}y^2+-a_{11}y+qa_{21}xy-qa_{22}xy^2  \nonumber \\
f_7(x,y) &=& +qb_{21}y^2-b_{11}y+qb_{21}xy^2-qb_{22}xy^2  \nonumber \\
f_8(x,y) &=& -b_{11}y^2-b_{10}y+b_{22}x-b_{20}xy+b_{21}xy^2-b_{12}.
\nonumber
\end{eqnarray}

The expression of the curvature components $eq(55)$ and the deformed field
strength $eqs(56)$ are formally the same as those obtained by Kerner $%
[2,15]. $ The functions $f_i(x,y)$ $i=0,...,8$ can be interpreted as a
direct consequence of the non-commutativity property of the space.

The covariant $n=3$ differential calculus constructed in section $3$ and $4$
respectively for $q$ $3^{rd}$ and $N^{th}-$root of unity can be seen as a
genralization of the case $n=2$. However, one cannot see $d^2=0$ as a
certain limit of $d^3=0$ case. In the next section, we remind the
differential calculus $d^2=0$.

\section{Differential calculus with nilpotency $n=2$ \ on a reduced quantum
plane.}

We recall that the exterior differential $"d"$ on the reduced quantum plane
satisfies usual properties $[6-9]$, namely

$\ \ \ i/$ \ \ linearity,

$i$ $i$ $/$ Nilpotency,

\begin{equation}
d^2=0.
\end{equation}

$i$ $i$ $i$ /Leibniz rule,

\begin{equation}
d(uv)=d(u)v+(-1)^nud(v),
\end{equation}

where

\begin{equation}
u\in \Omega ^n,v\in \Omega ^m\;\hbox{and},\, d(x)=dx,\, d(y)=dy,%
\, d1=0.
\end{equation}

The deformed differential calculus satisfies:

\begin{equation}
xdx=q^2dxx
\end{equation}

\begin{equation}
xdy=qdyx+(q^2-1)dxy
\end{equation}

\begin{equation}
ydx=qdxy
\end{equation}

\begin{equation}
ydy=q^2dyy
\end{equation}

\begin{equation}
dydx=-q^2dxdy
\end{equation}

\begin{equation}
(dx)^2=(dy)^2=0.
\end{equation}

So, the differential algebra $\Omega $ is generated by $x,$ $y,$ $dx$ and $%
dy $, $\Omega =\{x,y,dx,dy\}.$

Using the standard realization of the differential $"d"$:

\begin{equation}
d=\frac \partial {\partial x}dx+\frac \partial {\partial y}dy=\partial
_xdx+\partial _ydy,
\end{equation}

one can prove that:

\begin{equation}
\partial _xx=1+q^2x\partial _x+(q^2-1)y\partial _y
\end{equation}

\begin{equation}
\partial _yx=qx\partial _x
\end{equation}

\begin{equation}
\partial _xy=qy\partial _x
\end{equation}

\begin{equation}
\partial _{y}y=1+q^{2}y\partial _{y}.
\end{equation}

We apply this covariant differential calculus to study the related gauge
field theory on $M_3(C).$

\section{ Gauge field theory on M$_3$(C) as reduced quantum plane with d$^2$%
=0}

Similarly, the covariant differential is defined as in section $6$:

\begin{equation}
D\Phi (x,y)=d\Phi (x,y)+A(x,y)\Phi (x,y).
\end{equation}

The expression of the curvature is:

\begin{equation}
D^{2}\Phi (x,y)=(dA(x,y)+A(x,y)A(x,y))\Phi (x,y)=R\Phi (x,y).
\end{equation}

The differential realization of $"d"$ $eqs(67-71)$ allows to rewrite the
expression of the curvature $R$:

\begin{equation}
R=(\partial _xA_y(x,y)-q\partial _yA_x(x,y))dxdy+A_x(x,y)dxA_y(x,y)dy.
\end{equation}

Using the differential calculus ~$eqs(61-71)$ on the reduced quantum plane
and the expressions of $A_x(x,y)$, $A_y(x,y)$ $eqs(53,54)$ as formal power
series, it is easy to establish:

\[
R=[\partial _xA_y(x,y)-q\partial
_yA_x(x,y)+A_x(x,y)A_y(qx,q^2y)-qAy(x,y)A_x(q^2x,qy)+
\]

\begin{equation}
(1-q)A_y(x,y)\{-qb_{12}-b_{10}y+q^2b_{22}x-q^2b_{11}y^2+qb_{20}xy+b_{21}xy^2%
\}]dxdy,
\end{equation}
this permit us to define
\begin{eqnarray}
F_{xy}^q &=&\partial _xA_y(x,y)-q\partial _yA_x(x,y)+%
A_x(x,y)A_y(qx,q^2y)-qAy(x,y)A_x(q^2x,qy)  \nonumber \\
&=&-q\{\partial _yA_x(x,y)-q^2\partial _xA_y(x,y)+%
A_x(q^2x,qy)A_y(x,y)-q^2A_x(x,y)A_y(q^2x,qy)\}  \nonumber \\
&=&-qF_{yx}^{q^2},
\end{eqnarray}
which is the $\ q$-deformed \ antisymetric field strengh.

The comparison of the two expressions of curvature $(d^3=0$ section $5$ and $%
d^2=0)$ will be given in the following section.

\section{Discussions and concluding remarks}

In this paper, we have constructed a differential calculus $n=3$ nilpotent
on the reduced quantum plane by mixing Kerner's idea and Coquereaux's
technics. The notion of covariance for this differential calculus is also
given and we show that there is a quantum group structure behind this
covariance. As an application, we have constructed a gauge theory based on
this calculus.

In the case $n=3,$ the expressions of curvature contain additionnal terms $%
eqs(55,57)$ compared with $eq(75).$ These terms can be interpreted as a
generic consequence of the extension of the differential calculus $d^2=0$ to
the higher order $d^3=0.$

We can also compare our results with those of $Kerner$ $\&$ $al$ $[2,15].$
In fact, $eqs(55,56)$ are formally the same as in $[2,15]$, they differ only
by the appearance of the deformation parameter $q$. However, there is no
analogous of $eq(57)$ in $[2,15]$. It is a direct consequence of the
noncommutativity of the space considered here.

In a forthcoming paper, we shall treat in a mathematical way \ the
correspondance between this calculus and the Heisenberg algebra. This
correpondance is based on the bargman Fock reprensentation and will give a
new oscillator algebra. To study the minimization of incertitude principal
in this case, we will try to find the eigenvectors of the annihilation
operator in the way to construct the corresponding \ Klauder s coherent
states $\left[ 13\right] $.

\vspace{0.5cm}

\section*{Acknowledgments}

The authors Y. Hassouni and E. H. Zakkari wish to thank the Abdus Salam
International Centre for Theoretical Physics for financial and scientific
support, where a great part of this work has been done. This work was done
within the framework of Associateship and Federation Arrangement  Schemes of
the Abdus Salam ICTP.\textbf{\ }

\newpage

\section*{Appendix:}

We start by writting a priori $xdx,$ $xdy$ $ydx$ and $ydy$ in terms of $dxx,$
$dyx,$ $dxy$ and $dyy$ ie:

\begin{equation}
xdx=a_1dxx+b_1dyx+c_1dxy+d_1dyy
\end{equation}

\begin{equation}
xdy=a_2dxx+b_2dyx+c_2dxy+d_2dyy
\end{equation}

\begin{equation}
ydx=a_3dxx+b_3dyx+c_3dxy+d_3dyy
\end{equation}

\begin{equation}
ydy=a_4dxx+b_4dyx+c_4dxy+d_4dyy .
\end{equation}

Differentiating the commutation relation$\ xy=qyx$ and replacing $xdx$ and $%
xdy$ by their expressions in the formulae above, permit us to fix three
unknown coeficients. Actually, we have $nine$ independant parameters.

The left coaction of \ $F$ on a quantum plane is defined by:

\[
x_{1}=a\otimes x+b\otimes y
\]

\[
y_1=c\otimes x+d\otimes y.
\]

Hence

\[
dx_{1}=a\otimes dx+b\otimes dy
\]

\[
dy_{1}=c\otimes dx+d\otimes dy.
\]

We impose that the relations between $x_1,$ $y_1$ and \ $dx_1,$ $dy_1$ be
the same as the relations betwen $x,$ $y$ and $\ dx,$ $dy;$ these conditions
yields to:

\[
a_{2}=\, a_{3}=\; a_{4}=\, %
b_{1}=b_{4}=c_{1}=c_{4}=d_{1}=d_{2}=d_{3}=0\, \hbox{and }d_{1}=\,
a_{4},
\]

So, the unknown coeficients $b_2,$ $b_3,$ $c_2,$ and $c_3$ can be expressed
in the terms of one unknown coeficient $a_1$. Indeed:

\[
b_{2}=\frac{q(1+a_{1})}{1+q^{2}}\ \ \ \ \ \ \ \ \ \ \ \ \ \ \ \ \ \ \ \ \ \
\ \ \ c_{2}=\ \frac{a_{1}q^{2}-1}{1+q^{2}}\
\]

\[
b_{3}=\frac{a_{1}-q^{2}}{1+q^{2}}\ \ \ \ \ \ \ \ \ \ \ \ \ \ \ \ \ \ \ \ \ \
\ \ \ \ \ c_{3}=\frac{q(1+a_{1})}{1+q^{2}}.
\]

Differentiating the relations $(74-77)$ and noticing that $\ dxdx$, $d^2x$, $%
dydy$ and $d^2y$ are independant, we find \ $a_1=q^2$. The left covariant
differential calculus on a reduced quantum plane is hence constructed $%
eqs(9-22).$

\newpage

\textbf{References:}

$\left[ 1\right] $- V. Abramov and R. Kerner, J. Math. Phys, \textbf{41},
(2000) 5598.

$\left[ 2\right] $ - V. Abramov and R. Kerner, hep-th/9607143 (1996).

$\left[ 3\right] $ - R. Kerner, Math-ph/0011023 (2000).

$\left[ 4\right] $- M. Daoud and Y. Hassouni, Int. Jour. Theor. Phys.
\textbf{36}, (1996) 37.

$\left[ 5\right] $- A. Elhassouni, Y. Hassouni and E. H. Tahri, Int. Jour.
Theor. Phys. \textbf{35 } (1996) 2517.

$\left[ 6\right] $-R.Coquereaux, A. O. Garcia and R. Trinchero,
Math-ph/9807012 (1998).

$\left[ 7\right] $- R. Coquereaux, A. O. Garcia and R. Trinchero, Phys.
Lett. B. \textbf{443} (1998) 221.

$\left[ 8\right] $- T. Brezinski and J. Rembielinski, J.\ Phys. \textbf{A25}
(1992) 1945.

$\left[ 9\right] $-J. Wess and B. Zumino, Nucl. Phys. B (Proc Suppl) \textbf{%
18} B (1990) 302.

$[10]$- H. Weyl, \textit{The thory of groups and quantum mechanics} (Dover
Publications, 1931).

$\left[ 11\right] $- R. Coquereaux, A. O. Garcia and R. Trinchero,
math-hep/9811017 (1998).

$\left[ 12\right] $- R. Coquereaux and G. E. Schieber, \textit{Action of
finite quantum group on the algebra of complex }$N\times N$\textit{\ matrices%
}. A.I.P Conference Proceedings.\textbf{45}.

$\left[ 13\right] $- \ Y. Hassouni and E. H . Zakkari, in preparation.

$\left[ 14\right] $- A. K. Mishra and G. Rajasekaran, J. Math. Phys. \textbf{%
38} (1997) 1.

$[15]$-R. Kerner,\textit{\ }math-ph/0004031 (2000).

$\left[ 16\right] $- J. Madore, S. Schraml and J. Wess, Eur. phys. J . C
\textbf{16} (2000) 161.

$\left[ 17\right] $-B. Jurcco, P. Schupp and J. Wess, Nucl. phys. B. \textbf{%
584}. (2000). 784.

$\left[ 18\right] $-B. Jurcco, S. Schraml, P. Schupp and J. Wess, Eur. phys.
J C \textbf{17} (2000) 521.

$\left[ 19\right] $-B. Jurcco, P. Schupp and J. Wess, Mod. Phys. Lett. A.
\textbf{16}. (2001) 343.$\ $

\end{document}